\documentclass[twocolumn, amsmath,amssymb,aps, superscriptaddress]{revtex4-1}

\usepackage{verbatim}
\usepackage{graphicx}
\usepackage{ifpdf} 
\usepackage{dcolumn}  
\usepackage{epsfig}
\usepackage{color}
\usepackage[usenames,dvipsnames]{xcolor}
\usepackage{siunitx}

\newcommand{\kT}{k_\mathrm{B}T}
\newcommand{\lgc}{\ell_{\rm GC}}
\newcommand{\lb}{\ell_{\rm B}}
\newcommand{\kkkk}{K$_4$Fe(CN)$_6$}
\newcommand{\celsius}{\ensuremath{^\circ}C}
\def\rmd{{\mathrm{d}}}
\def\eq{Eq.}
\def\eqs{Eqs.}
\def\ie{{\sl i.e.}}
 
\begin{document}

\title{Experimental Evidence for Algebraic Double-Layer Forces}

\author{Biljana Stojimirovi\'c}
\affiliation{Department of Inorganic and Analytical Chemistry, University of Geneva,
Sciences II, 30 Quai Ernest-Ansermet, 1205 Geneva, Switzerland}

\author{Mark Vis}
\email{E-mail: \texttt{m.vis@tue.nl}}

\affiliation{Laboratory of Physical Chemistry, Faculty of Chemical Engineering and Chemistry \& Institute for Complex Molecular Systems, Eindhoven University of Technology, PO Box 513, Eindhoven 5600 MB, The Netherlands}

\author{Remco Tuinier}
\affiliation{Laboratory of Physical Chemistry, Faculty of Chemical Engineering and Chemistry \& Institute for Complex Molecular Systems, Eindhoven University of Technology, PO Box 513, Eindhoven 5600 MB, The Netherlands}
\affiliation{Van 't Hoff Laboratory for Physical and Colloid Chemistry, Debye Institute for Nanomaterials Science, Utrecht University, Padualaan 8, Utrecht 3584 CH, The Netherlands}

\author{Albert P. Philipse}
\affiliation{Van 't Hoff Laboratory for Physical and Colloid Chemistry, Debye Institute for Nanomaterials Science, Utrecht University, Padualaan 8, Utrecht 3584 CH, The Netherlands}

\author{Gregor Trefalt}
\email{E-mail: \texttt{gregor.trefalt@unige.ch}}
\affiliation{Department of Inorganic and Analytical Chemistry, University of Geneva,
Sciences II, 30 Quai Ernest-Ansermet, 1205 Geneva, Switzerland}

\date{\today}

\begin{abstract}
According to conventional wisdom electric double-layer forces normally decay exponentially with separation distance. Here we present experimental evidence of algebraically decaying double-layer interactions. We show that algebraic interactions arise in both strongly overlapping as well as counterion-only regimes, albeit the evidence is less clear for the former regime. In both of these cases the disjoining pressure profile assumes an inverse square distance dependence. At small separation distances another algebraic regime is recovered. In this regime the pressure decays as the inverse of separation distance.
\end{abstract}

\maketitle


\section{Introduction}

The repulsion between two charged surfaces immersed in electrolyte solution is usually evaluated under the assumption that electrical double-layers only weakly overlap~\cite{Derjaguin1940, Derjaguin1941, Verwey1948, Kruyt1952, Evans1999, Russel1989, Lyklema2005a, Israelachvili2011, Elimelech2013}. The implication is that the electrical potential in the mid-plane between the two surfaces is small, which allows for an approximate solution of the Poisson--Boltzmann (PB) equation, that eventually leads to exponentially decaying double-layer repulsions, with a decay distance set by the Debye screening length. Together with van der Waals attractions, exponentially screened repulsions form the classical Derjaguin--Landau--Verwey--Overbeek (DLVO) potential \cite{Derjaguin1940, Derjaguin1941, Verwey1948, Kruyt1952, Evans1999, Russel1989, Lyklema2005a, Israelachvili2011, Elimelech2013}.
Usually exponential screening is assumed to be a generic feature of double-layer repulsions. However, when charged surfaces immersed in salt solutions approach each other within distances smaller than the Debye length, screening of charges diminishes, or even vanishes, and there is no reason to assume that under this condition exponential decay of double-layer repulsions will remain intact.

For long time, the general opinion about strongly overlapping double-layers in salt solutions was that one has to resort to numerical solutions of the PB-equation to obtain an interaction potential for which there are no simple expressions \cite{Derjaguin1940, Derjaguin1941, Verwey1948, Kruyt1952, Evans1999, Russel1989, Lyklema2005a, Israelachvili2011, Elimelech2013}. Recently, however, it was shown that the weak-potential Debye--H\"uckel (DH) limit for surfaces far apart, has a pendant limit in the form of a weak electric field for surfaces in close proximity \cite{Philipse2013}. In the DH case, ions diffuse in a weak but spatially varying potential, whereas in the pendant situation ions roam around in a possibly high but spatially constant electrical potential, a case also known as the Donnan equilibrium \cite{Philipse2011}. In the zero-field Donnan limit the repulsion between two surfaces in a salt solution can be calculated analytically \cite{Philipse2013,Philipse2017,Vis2018}. The result is an algebraic repulsion where the disjoining pressure decays as an inverse square of separation distance. Strongly overlapping double-layers were further studied for surfaces of finite size and non-uniformly charged surfaces. In these situations charge overspill effects additionally modifies the algebraically decaying pressure~\cite{Ghosal2016, Ghosal2017}.

Algebraic dependencies are also found in counterion-only systems. These are systems, where charged surfaces are in a bath of only counterions, namely if the surface is negatively charged the adjacent fluid only contains cations. In such systems the counterion concentration profile adjacent to the single charged plate decays algebraically \cite{Gouy1910, Chapman1913, Markovich2016a}, in contrast to the system with salt where the counterion profile decays exponentially. The solution for the interaction between two plates immersed in a counterion-only bath was first found by Langmuir \cite{Langmuir1938}. He calculated that in the limit of high charge on the plates the pressure profile also has an algebraic dependence of $1/h^2$. Similar power-law dependencies were also observed when investigating ion-ion correlation effects~\cite{Kjellander2009, Attard1988a}.

While reports of experimentally measured exponential double-layer forces are abundant~\cite{Finlayson2016,Sainis2008,Smith2018a,Israelachvili1978,Prieve1999,Popa2010c,Smith2016}, there have been only a few articles showing algebraic double-layer forces \cite{Briscoe2002a, MontesRuiz-Cabello2015a, Moazzami-Gudarzi2016a} in literature. In all the presented cases these forces were measured in counterion-only regimes. On the other hand algebraic interactions for the strongly overlapping double-layers have only been studied theoretically~\cite{Philipse2013,Philipse2017,Vis2018}, and experimental instances of algebraic forces between surfaces in 1:1 salt solutions, as far as we know, have not been reported yet. The main aim of this paper is to present direct and quantitative evidence for the existence of algebraic interactions between surfaces immersed in simple salt solutions. We also explore different regimes where the algebraic forces are important experimentally.


\section{Materials and methods}
\subsection{Force Measurements}
Forces were measured between spherical silica particles (Bangs Laboratories Inc, USA) with an average reported size of 5.2 $\mu$m. The colloidal probe technique based on atomic force microscopy was used~\cite{Borkovec2012,Butt2005,Butt1991,Ducker1991}. A single silica particle was glued on tip-less cantilever (MicroMasch, Tallin, Estonia) by first immersing the cantilever in a small drop of glue (Araldite 2000+). The substrate was made separately by spreading silica particles on a quartz microscope slide (Plano GmbH, Wetzlar, Germany), which was previously cleaned in piranha solution (3:1 mixture of H$_2$SO$_4$ (98\%) and H$_2$O$_2$ (30\%)). Cantilevers with glued-on particles and substrate were both heated at 1200\celsius\ for 2~h to achieve firm attachment and removal of the glue. During this sintering process, particles shrink about 15\%, so the average diameter is 4.4 $\mu$m~\cite{Uzelac2017}. A root mean square (RMS) roughness of 0.63~nm was measured by AFM imaging in liquid. Solutions were made using KCl (Sigma Aldrich) and Milli-Q water (Millipore). Solutions of pH 10 were made with addition of 1M KOH (Acros Organics), and solutions of pH 3 with 1M HCl (Fisher Scientific). For experiments at pH 5.6 no adjustment was done.

\begin{figure}[t]
\centering
\includegraphics[width=8.5cm]{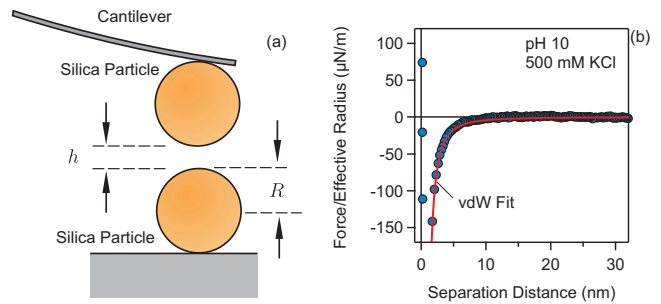}
\caption{(a) Schematic representation of colloidal probe experiment. Van der Waals force measured in 500~mM KCl. The extracted Hamaker constant is $H = 2.6\cdot 10^{-21}$~J \cite{Uzelac2017}.}
\label{fig:colloidal_probe}
\end{figure}

Force measurements were done at room temperature $23\pm 2$\celsius\ with a closed loop AFM (MFP-3D, Asylum Research) mounted on an inverted optical microscope (Olympus IX70). Both cantilever and the substrate were cleaned in ethanol and water, and plasma treated for 20 minutes. The substrate with particles was mounted on the fluid cell. The geometry of the experiment is shown in Fig.~\ref{fig:colloidal_probe}a. The deflection of the cantilever was recorded when the particle on the cantilever was centered above the selected one on the substrate with the precision of about 100~nm. For one pair of particles, the deflection is recorded in 150 approach-retract cycles with the cantilever velocity of 400~nm/s. The measurement was done on 3--5 different pairs for each solution concentration. The spring constant of the cantilever was determined by the Sader method~\cite{Sader1999}, and the deflection was converted to force using Hooke’s law. The approach part of the recorded curves is averaged and down-sampled for increasing the force resolution and readability of the figures. These curves are then also used for theoretical analysis.

Our analysis of the experimental forces only focuses on the electrostatic double-layer contribution. However in some cases the van der Waals forces become non-negligible. Therefore the van der Waals force for the present silica--silica system was measured in 500~mM KCl and the respective Hamaker constant was extracted, see Fig.~\ref{fig:colloidal_probe}b. In order to be sure that we are measuring only the van der Waals force and the electrostatic interactions are completely screened at 500~mM, we have also preformed measurements in 1 M KCl. The 500~mM and 1~M curves overlap, and therefore we are confident that the electrostatic forces are completely screened. The van der Waals contribution was then subtracted from all the experimental force profiles. The measured forces $F(h)$ between two spherical particles were then converted to the equivalent disjoining pressure between two plates, $\Pi(h)$, by calculating the derivative of the force versus the separation distance, $h$, and by applying the Derjaguin approximation~\cite{Russel1989}
\begin{equation}
\Pi(h) = -\frac{1}{\pi R}\cdot \frac{\rmd F(h)}{\rmd h} ,
\label{eq:force_to_press}
\end{equation}
where $R$ is the particle radius.

\subsection{Theory}

\subsubsection{Poisson--Boltzmann Theory}

The disjoining pressure between two charged plates is calculated by solving the Poisson--Boltzmann (PB) equation
\begin{equation}
\frac{\rmd^2 \Psi(x)}{\rmd x^2} = -\frac{\beta e_0^2}{\varepsilon\varepsilon_0}\sum_i c_i \,e^{-z_i \Psi(x)} ,
\label{eq:pb}
\end{equation} 
where $\Psi = \beta e_0 \psi$ is the rescaled dimensionless electrostatic potential, $x$ is the coordinate normal to the plates, $c_i$ is the bulk concentration of ion $i$, $z_i$ is the valence of ion $i$, $e_0$ is the elementary charge, $\varepsilon_0$ is the dielectric permittivity of vacuum, $\varepsilon$ is the relative dielectric permittivity of water and $\beta = 1/(\kT)$ is the inverse thermal energy, where $k_{\rm B}$ is the Boltzmann constant and $T$ is the temperature. Throughout, $T = 298$~K and $\varepsilon = 80$ are used as appropriate for dilute aqueous solutions. The PB equation is solved numerically with different boundary conditions. One typically assumes a constant charge (CC) or constant potential (CP) on the plates at different distances $h$, however the charge on the plates can be also regulated upon approach~\cite{Avni2019,Behrens1999,Kirkwood1952,Markovich2016,Adzic2014}. Here we employ constant regulation approximation, where we introduce the regulation parameter $p$ \cite{Behrens1999, Trefalt2016}. The regulation parameter is defined as
\begin{equation}
p = \frac{C_{\rm dl}}{C_{\rm dl} + C_{\rm in}} ,
\label{eq:regulation}
\end{equation}
where $C_{\rm dl}$ and $C_{\rm in}$ are double-layer and inner-layer capacitance, respectively. The regulation parameter interpolates between CP ($p = 0$) and CC ($p=1$) boundary conditions and can be used for describing charge-regulation in general way.

The solution of the PB equation yields the potential profile $\psi(x)$ between two charged plates positioned at $x = -h/2$ and $x = +h/2$. The potential at the mid-plane, $\psi_{\rm M} = \psi(0)$, then permits to calculate the disjoining pressure
\begin{equation}
\Pi (h) = \kT \sum_i c_i \left( e^{-z_i \Psi_{\rm M}(h)}-1 \right) ,
\label{eq:pressure}
\end{equation}
where $h$ is the distance between the plates. 

Further details of the implementation of the full PB theory including constant charge regulation model are given in Ref.~\citenum{Uzelac2017}.

\subsubsection{Strongly Overlapping Double-Layers}

For strongly overlapping double-layers analytical approximations of the PB equation can be found as was demonstrated recently~\cite{Philipse2013,Vis2018,Philipse2017}. These approximations are applicable for symmetric $z:z$ electrolytes and here we will focus only on the 1:1 case. For 1:1 electrolytes, the disjoining pressure can be calculated by simplifying \eq~(\ref{eq:pressure})
\begin{equation}
\Pi = 2\kT c_s [\cosh(\Psi_{\rm M}) - 1 ] .
\label{eq:pressure_11}
\end{equation}
For strongly overlapping double-layers, the electrical potential between the plates deviates only little from its average value, \ie\ the electric field in the gap is approximately zero. The mid-plane potential in the equation above can be then be replaced by a constant $\Psi_{\rm M} = \overline{\Psi}$, where $\overline{\Psi}$ turns out to be the dimensionless Donnan potential. The value of the Donnan potential can be determined from the electro-neutrality condition for the two plates and the intermediate salt solution
\begin{equation}
\frac{\overline{c_+} - \overline{c_-}} {2c_s} = \frac{\lambda}{h},
\label{eq:neutrality}
\end{equation}
where $\overline{c_+}$ and $\overline{c_-}$ refer to the cation and anion number concentrations between the plates, respectively. We have introduced characteristic length $\lambda$ defined as
\begin{equation}
\lambda = \frac{\sigma}{c_s e_0} = \frac{4}{\kappa^2 \lgc} ,
\label{eq:lambda}
\end{equation}
where $\sigma$ is the surface charge density of the plates and $\kappa = \sqrt{8\pi \lb c_s}$ is the inverse Debye length, with $\lb = \beta e_0^2 / (4\pi \varepsilon\varepsilon_0)$ the Bjerrum length, and 
\begin{equation}
\lgc = \frac{e_0}{2\pi \lb \sigma} 
\label{eq:lgc}
\end{equation}
is the Gouy--Chapman length. At separation distances below $\lambda$ the contribution of counterions to pressure is dominant, while at $h>\lambda$ background salt contribution becomes more important. The Gouy--Chapman length is connected to the thickness of the double-layer in the counerion-only case and it is an analog to Debye length in systems containing salt.

By combining \eq~(\ref{eq:neutrality}) with the Boltzmann equilibrium for exchanging ions between bulk and the gap
\begin{equation}
\overline{c_{\pm}} = c_s e^{\mp \overline{\Psi}} ,
\label{eq:boltzmann}
\end{equation}
we can calculate the value of the Donnan potential
\begin{equation}
\overline{\Psi} = -\sinh^{-1}\left( \frac{\lambda}{h} \right) .
\label{eq:donnan}
\end{equation}
The zero-field pressure can be finally calculated by inserting \eq~(\ref{eq:donnan}) into \eq~(\ref{eq:pressure_11})
\begin{equation}
\Pi = 2\kT c_s \left( \sqrt{1 + \left( \frac{\lambda}{h} \right)^2}-1 \right) .
\label{eq:zero-field1}
\end{equation}
For $\lambda/h \ll 1$, \ie\ large separation and low surface charge, the above equation reduces to
\begin{equation}
\Pi = \kT c_s \left( \frac{\lambda}{h} \right)^2 .
\label{eq:zero-field2}
\end{equation}
The pressure for strongly overlapping double-layers therefore decays algebraically $\Pi \propto 1/h^2$. Another algebraic dependence is recovered at small separations
\begin{equation}
\Pi = 2\kT c_s \cdot \frac{\lambda}{h} = \frac{2\sigma\kT}{e_0}\cdot\frac{1}{h}  .
\label{eq:ideal-gas}
\end{equation}
This latter equation is similar to the ideal-gas equation of state, $P = N\kT /V$, and we will refer to it as the ideal-gas equation~\cite{Markovich2016a}. Note that the ideal-gas analogy comes from the fact that the pressure between the plates in this regime can be calculated by the ideal-gas equation.

The disjoining pressure in the zero-field approximation is derived here for plates with constant charge densities. When one includes charge regulation effects, the resulting pressure decay is not affected, but only the amplitude weakens~\cite{Vis2018,Philipse2017}. It has been also shown that the Donnan potential in \eq~(\ref{eq:donnan}) results from the asymptotic solution of the PB-equation for vanishing electric field~\cite{Philipse2013}.

\subsubsection{Counterion-Only Double-Layers}

In the counterion only case the disjoining pressure also decays algebraically~\cite{Langmuir1938, MontesRuiz-Cabello2015a, Briscoe2002}. Here the only ionic species are the counterions to the surfaces. If one assumes a positively charged surface and anions as counterions, the PB equation reduces to
\begin{equation}
\frac{\rmd^2 \Psi(x)}{\rmd x^2} = \frac{\beta e_0^2c_-}{\varepsilon\varepsilon_0} e^{\Psi(x)} .
\label{eq:pb-counter}
\end{equation}
This equation can be solved analytically for two charged plates and the resulting potential profile is
\begin{equation}
\Psi(x) = \Psi_{\rm M}-\ln \left[ \cos^2  \left( \frac{x\gamma}{\lgc} \right) \right] ,
\label{eq:potential-counter}
\end{equation} 
where $\gamma = e^{\Psi_{\rm M} - \Psi_{\rm D}}$, $\Psi_{\rm D}$ being the diffuse-layer potential at the isolated plate surface. The disjoining pressure in this case is equal to the pressure between the plates, since the bulk pressure vanishes without salt and is calculated by using \eq~(\ref{eq:potential-counter})
\begin{equation}
\Pi = \kT c_- e^{\Psi_{\rm M}} = \frac{2\varepsilon\varepsilon_0}{\beta^2 e_0^2 \lgc^2} \gamma^2 .
\label{eq:pressure-counter}
\end{equation}
The equation above shows that $\gamma^2$ is actually the rescaled pressure. For the constant charge boundary conditions the relation between the separation distance and pressure becomes~\cite{MontesRuiz-Cabello2015a}
\begin{equation}
h = \frac{2\lgc}{\gamma} \arctan \left(\gamma^{-1} \right) .
\label{eq:pressure-counter-cc}
\end{equation}
At large separation distances the pressure tends to zero and one can approximate $\arctan(x) \approx \frac{\pi}{2}$. With this we can rewrite the equation above as
\begin{equation}
\Pi = \frac{\pi}{2\beta\lb}\cdot\frac{1}{h^2} .
\label{eq:pressure-lang}
\end{equation}
Here, the decay of the pressure is also algebraic. This equation was first derived by Langmuir~\cite{Langmuir1938} and we will refer to it as the Langmuir equation. Note that the Langmuir equation features the same $1/h^2$ dependence as the zero-field result given in \eq~(\ref{eq:zero-field2}), albeit the pre-factor and physical origin are different.

In the opposite limit, \ie\ short separation distances, where $\arctan (x) \approx x$,  \eq~(\ref{eq:pressure-counter-cc}) reduces to the ideal-gas equation shown in \eq~(\ref{eq:ideal-gas}). Therefore both the counterion-only and zero-field approximation lead to the same result for small separation distances. The ideal-gas equation can be understood for the counterion-only case in the following way. When plates come sufficiently close together, all the coions are expelled and only counterions remain. Due to charge neutrality, the number of counterions in the gap must be equal to the number of charges on the plates. The distance between the plates and the area of the plates therefore determine the concentration of the ions in the gap as $c = 2\sigma/(e_0 h)$; multiplying this concentration by $\kT$ leads to the disjoining pressure in \eq~(\ref{eq:ideal-gas}).

For the case of asymmetric $z:1$ salts, where $z$ is the valence of the coion, the multivalent coions get expelled from the gap between the charged plates upon approach \cite{MontesRuiz-Cabello2015a, Uzelac2017}. This situation results in a counterion-only system, however the salt is still present in the bulk. We can correct the Langmuir pressure for the bulk contribution. If we add an additional term in the $\arctan(x) \approx \frac{\pi}{2} - \frac{1}{x}$ expansion we arrive at~\cite{Moazzami-Gudarzi2016a}
\begin{equation}
\Pi = \frac{\pi}{2\beta\lb}\cdot\frac{1}{(h+2\lgc)^2}-\frac{(z+1)c_s}{\beta} .
\label{eq:pressure-lang-corr}
\end{equation}
We refer to this approximation as the corrected Langmuir equation.

\section{Results and Discussion}

We use the colloidal probe technique based on AFM microscopy to experimentally study algebraic double-layer forces between colloidal particles. Specifically, we measure the double-layer interactions between micron-sized silica particles in the presence of KCl at different pH. We further analyze the forces in \kkkk\ to show experimental evidence of counterion-only induced algebraic double-layer interactions.

\begin{figure}[t]
\centering
\includegraphics[width=8.5cm]{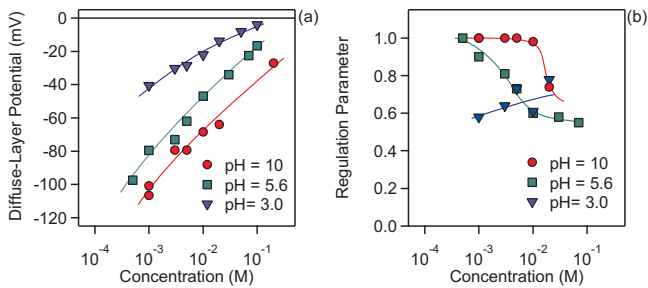}
\caption{(a) Diffuse-layer potential and (b) regulation parameter as a function of KCl concentration for different pH conditions. The diffuse-layer potentials and regulation parameters were extracted by fitting experimental force data with full PB theory~\eq~(\ref{eq:pb}) using constant regulation approximation. Further details on this fitting procedure can be found in Ref.~\citenum{Uzelac2017}.}
\label{fig:potential-regulation}
\end{figure}

First the van der Waals forces are subtracted from all the experimentally measured forces, which yields only the double-layer contribution to the force profile. The resulting double-layer forces are then converted to disjoining pressures between plates by calculating the derivative of the force versus the separation distance. These experimental pressure profiles are then fitted with the full PB theory using the constant regulation approximation and the silica surface properties are extracted from the fits. Note that the concentrations are fixed to nominal values during fitting. The diffuse-layer potentials and regulation parameters for silica particles extracted from this fitting procedure are shown in Fig.~\ref{fig:potential-regulation}. 
As expected for silica surfaces in aqueous solutions, the double-layer potentials are increasing with increasing salt concentration due to electrostatic screening. The particles acquire most negative charge at pH 10, where the fraction of charged silanol groups is the highest. On the other hand, regulation parameters are decreasing with increasing concentration at pH 10 and pH 5.6, while they are modestly increasing at pH 3. These observations are consistent with earlier reports, where a similar decrease in the regulation parameters was reported for pH 10, while at pH 4 the regulation parameters were constant~\cite{Smith2018a}. Note that regulation parameter of 1 represents constant charge conditions (CC), while at constant potential (CP) conditions the regulation parameter is 0 and the surface charge varies strongly when the two surfaces approach each other. Therefore at high pH and low concentration the silica particles behave as CC surfaces, while they regulate more strongly at low pH and high salt concentrations. Surface properties of silica reported in Fig.~\ref{fig:potential-regulation} are not the main interest of the present paper, but they are used to determine which asymptotic regimes are applicable at certain solution conditions. The double-layer potentials are further converted by employing the Graham equation to surface charge density and used to calculate characteristic length scales, such as Gouy--Chapman length.

\begin{figure}[t]
\centering
\includegraphics[width=8.5cm]{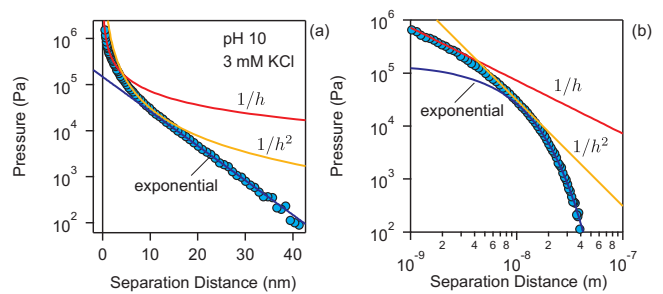}
\caption{Comparison of (a) log--lin and (b) log--log representations for disjoining pressure between two silica particles measured at pH 10 and 3 mM KCl. The algebraic $1/h$ and $1/h^2$, as well as exponential curves are also shown.}
\label{fig:force-example}
\end{figure}

The main focus of the present paper is the analysis of the asymptotic laws in the experimental double-layer interactions. In Fig.~\ref{fig:force-example}, an example of disjoining pressure measured at pH 10 and 3 mM KCl is shown. 
The exponential function $e^{-\kappa h}$, and two algebraic functions $1/h$ and $1/h^2$ are also shown in the figure. Note that the amplitudes of these general functions are adjusted to fit the experimental data in order to show at which separation range such dependencies are applicable. The data is shown in log--lin and log--log representations for clarity. This simple comparison reveals that the experimental data has an exponential dependence only at separation distances above $\sim 15$~nm, where the double-layers from the two surfaces are only weakly overlapping. At intermediate distances between 5 and 15~nm the double-layers overlap strongly and the $1/h^2$ dependence becomes applicable, while at distances below 5~nm the pressure decreases as $1/h$.

\begin{figure*}[t]
\centering
\includegraphics[width=15cm]{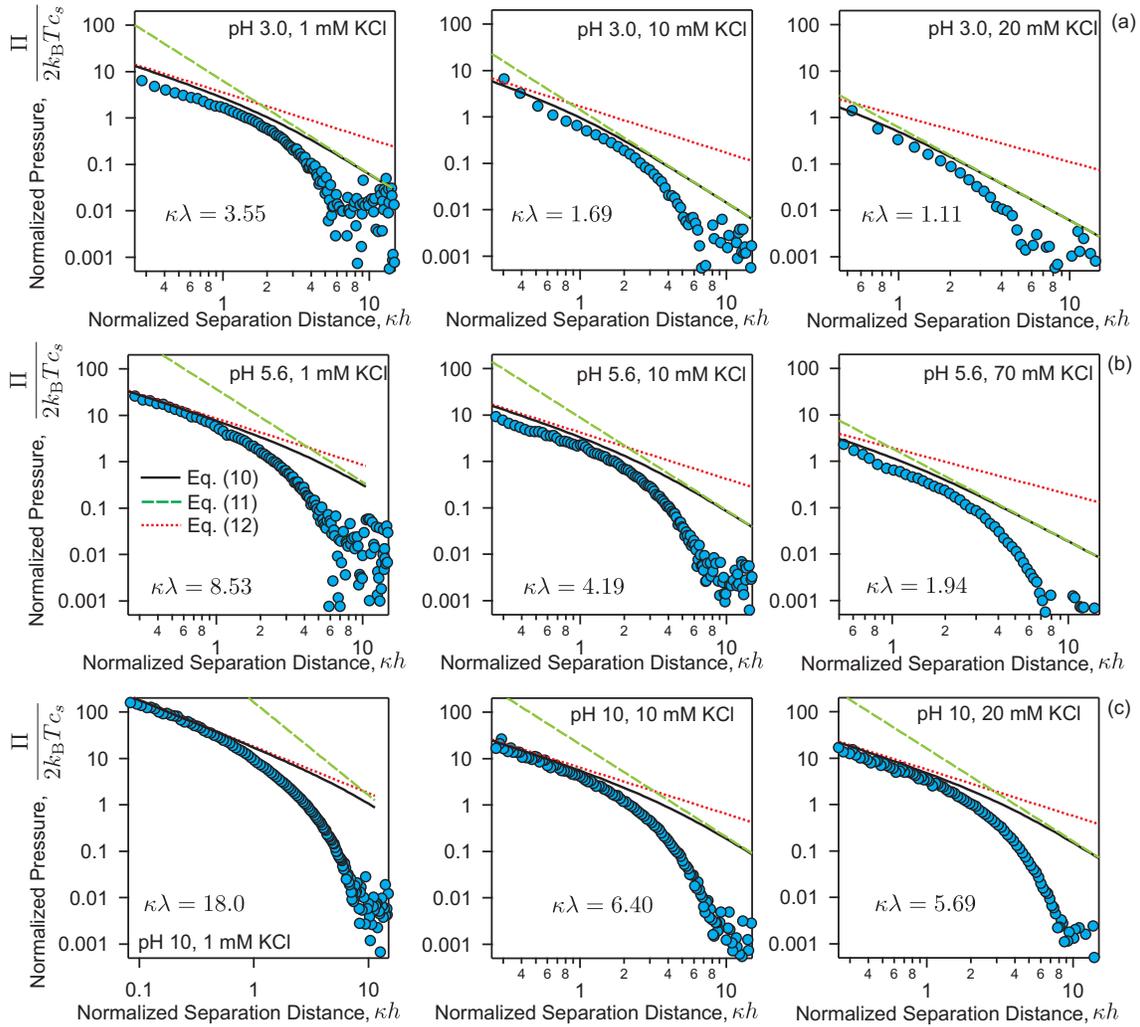}
\caption{Zero-Field theory compared to experimental disjoining pressures measured at (a) pH 3.0, (b) pH 5.6, and (c) pH 10. Different columns present different salt levels. Experimental data is shown with symbols, while full, dashed, and dotted lines represent \eqs~(\ref{eq:zero-field1}), (\ref{eq:zero-field2}), and (\ref{eq:ideal-gas}), respectively.}
\label{fig:force-zero-field}
\end{figure*}

Let us now have a more detailed look at the applicability of the approximations introduced in the theory section. First we will focus on strongly overlapping double-layers, where the zero-field approximation shown in \eq~(\ref{eq:zero-field1}) is expected. In Fig.~\ref{fig:force-zero-field}, the disjoining pressures for different pH conditions and different salt levels are shown. Note that pressures are normalized to bulk osmotic pressure while distance is normalized with Debye length.
Together with experimental data we plot the algebraic zero-field approximations shown in \eqs~(\ref{eq:zero-field1}), (\ref{eq:zero-field2}), and (\ref{eq:ideal-gas}). We further denote the value of $\kappa \lambda$ in the sub-figures, which represents the value of the characteristic length in comparison to the Debye length and permits the evaluation of the accuracy of the zero-field approximations~\cite{Philipse2013}.

In the top row of Fig.~\ref{fig:force-zero-field}, the data for pH 3 are shown. At pH 3 the silica surface exhibits the lowest surface charge and the regulation parameter is around 1/2 as evident from Fig.~\ref{fig:potential-regulation}. By comparing the experimental pressures to the calculated ones, one can deduce that the double-layers start to overlap strongly at distances shorter than about 2-3~$\kappa h$ and the zero-field approximation \eq~(\ref{eq:zero-field1}) becomes accurate. The simplified inverse square separation dependence $1/h^2$ shown in \eq~(\ref{eq:zero-field2}) represents a good approximation at intermediate separations at about $\kappa h \approx 1$ and gets more accurate at higher salt concentrations. The accuracy of the $1/h^2$ dependence, \eq~(\ref{eq:zero-field2}), is good at $\kappa \lambda \le 1$, as discussed earlier~\cite{Philipse2013}, therefore one needs to go to high salt concentrations and low pH for it to become applicable. Furthermore for pH 3 solutions the ideal-gas equation is not accurate since the surfaces strongly regulate and the ideal-gas equation is derived for constant charge surfaces.

At intermediate pH values of 5.6 shown in Fig.~\ref{fig:force-zero-field}b, the zero-field approximation works the best at low concentration and small separation distances. The simple $1/h^2$ dependence, \eq~(\ref{eq:zero-field2}), gets again more accurate at higher concentrations, but it is less accurate compared to pH 3 case, because the magnitude of the charge on silica is higher at higher pH. The ideal-gas equation works best at low concentrations, since in this case the regulation parameter is close to 1 and the surface has  approximately constant charge.

When pH is increased to 10 (Fig.~\ref{fig:force-zero-field}c), the $1/h^2$ approximation gets even less accurate since the magnitude of the charge again increases and consequently $\kappa \lambda$ further increases. On the other hand, the particles now follow the CC behavior and the ideal-gas $1/h$ dependence is highly accurate at small separations.

Looking at all the sub-figures in Fig.~\ref{fig:force-zero-field}, one can deduce that the simple algebraic $1/h^2$ dependence from \eq~(\ref{eq:zero-field2}) describes silica--silica interactions at intermediate separation distances at low pH and high salt concentrations, where $\kappa \lambda \sim 1$. Note that $\kappa \lambda \ll 1$ is not reached experimentally and therefore the agreement between \eq~(\ref{eq:zero-field2}) and experimental curves is not perfect. The ideal-gas relation \eq~(\ref{eq:ideal-gas}) is accurate at small separation distances, typically below $\kappa h \le 1$, and at low concentrations and high pH, where the regulation parameter is close to 1. Therefore experimental conditions where double-layers overlap strongly and algebraic interactions can be reached in 1:1 electrolytes.

\begin{figure}[t]
\centering
\includegraphics[width=8.5cm]{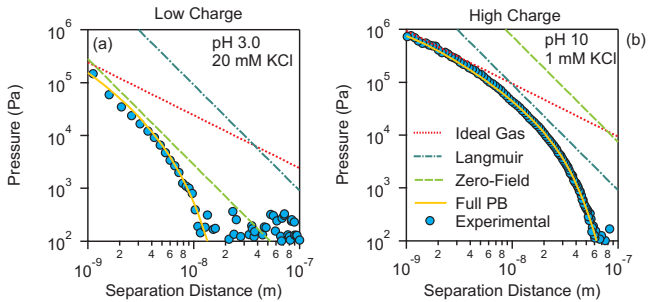}
\caption{Comparison of zero-field and Langmuir approximation for the (a) low charge and (b) high charge experimental cases.}
\label{fig:force-counerion}
\end{figure}

Let us now address the comparison of the zero-field approximation \eq~(\ref{eq:zero-field2}) with Langmuir expression, \eq~(\ref{eq:pressure-lang}). While the former is valid for strongly overlapping double-layers in 1:1 electrolyte, the latter is derived for counterion-only double-layers. Both of these approximations are valid at the intermediate separation distances and they both feature an inverse square distance dependence $1/h^2$. The difference between them is in the amplitude, \ie\ strength of the interaction. In Fig.~\ref{fig:force-counerion} a comparison of the algebraic interactions is made for weakly and highly charged systems, respectively.
The left sub-figure shows the disjoining pressure at pH~3 and 20~mM KCl, while the right subfigure shows the interaction at pH~10 and 1~mM KCl. In both cases the full PB with constant regulation approximation is used to fit the experimental data, and the PB theory perfectly describes the data in the whole distance range. At pH 10 the experimental pressure follows the $1/h$ ideal-gas dependence at small separations, while at pH 3 this regime is not recovered due to stronger charge regulation of silica surfaces at this condition. The zero-field $1/h^2$ dependence, \eq~(\ref{eq:zero-field2}), predicts the behavior rather well at pH 3, while it largely overestimates the repulsion at pH 10. The opposite is true for the Langmuir equation, which works better at pH 10 and it overestimates the pressure at pH 3. These results are consistent with the range of validity for both models. While the zero-field approximation is valid for weakly charged surfaces, the Langmuir equation is applicable for highly charged surfaces~\cite{Markovich2016a, Philipse2013, Philipse2017}.

\begin{figure}[t]
\centering
\includegraphics[width=8.5cm]{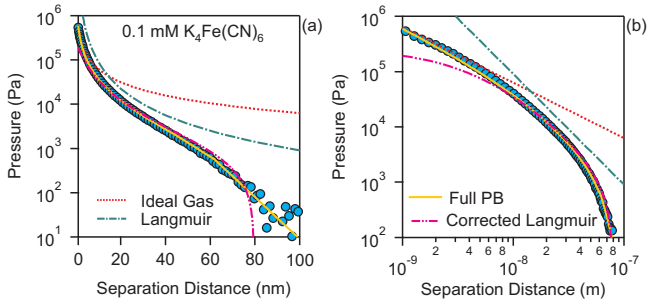}
\caption{Experimentally measured disjoining pressure in the presence of 0.1~mM \kkkk\ at pH 10. (a) Log--lin and (b) lin--lin representations of experimental data. The full PB model for the mixture of $1$:$z$ and 1:1 electrolyte is shown together with ideal-gas (\ref{eq:ideal-gas}), Langmuir (\ref{eq:pressure-lang}), and corrected Langmuir (\ref{eq:pressure-lang-corr}) approximations.}
\label{fig:force-coions}
\end{figure}

Another important experimental case where algebraic interactions are important is the multivalent coion case. In these systems, the multivalent ions have the same sign of charge as the surface. When two charged plates approach each other in these salts, at a certain distance, the multivalent coions are expelled from the gap between the charged plates \cite{Moazzami-Gudarzi2016a, MontesRuiz-Cabello2015a, Uzelac2017}. After the expulsion of the coions, the systems behaves as counterion-only and the pressure decays algebraically. An example of such interactions between two silica colloids in the presence of \kkkk\ salt at pH 10 is shown in Fig.~\ref{fig:force-coions}.
Again the disjoining pressure profile can be perfectly described with the full PB model with constant regulation approximation. At separation distances above $\sim 80$~nm, the pressure decays exponentially, and here both coions and counterions are still present in the gap. By lowering the separation distance, the experimental curve starts to deviate from the exponential decay and this deviation marks the transition into the counterion-only regime. At distances below $\sim 70$~nm the four-valent [Fe(CN)$_6$]$^{4-}$ ions get expelled from the gap, and double-layer is comprised of only K$^+$ ions. Here the pressure decays algebraically. The Langmuir equation captures this $1/h^2$ dependence, but it overestimates the magnitude of the force. If one corrects the Langmuir equation by accounting for the ions in the bulk and expanding the approximation with one more term, the resulting \eq~(\ref{eq:pressure-lang-corr}) describes the data very well in the separation range from 5 to 80~nm. At distances below 10~nm, the ideal-gas equation offers a quantitative description of the data. In the multivalent coion systems where the pressure decays algebraically, the counterion-only regime is typically important in a very wide separation range, and the exponential double-layer forces become evident only at very large distances where they are weak. In the presence of multivalent coions, the aggregation of colloids is also driven by the algebraic forces and leads to the {\sl inverse} Schultze-Hardy rule reported recently by some of us \cite{Cao2015}. Similar algebraically decaying interactions were recently measured in the presence of like-charged polyelectrolytes \cite{Moazzami-Gudarzi2016a} and between stacked charged bilayers \cite{Hishida2017}. In the case of polyelectrolytes, the transition between algebraic and exponential regimes is even more abrupt, due to the larger charge of the polyelectrolytes as compared to multivalent coions.

\begin{figure}[t]
\centering
\includegraphics[width=8.0cm]{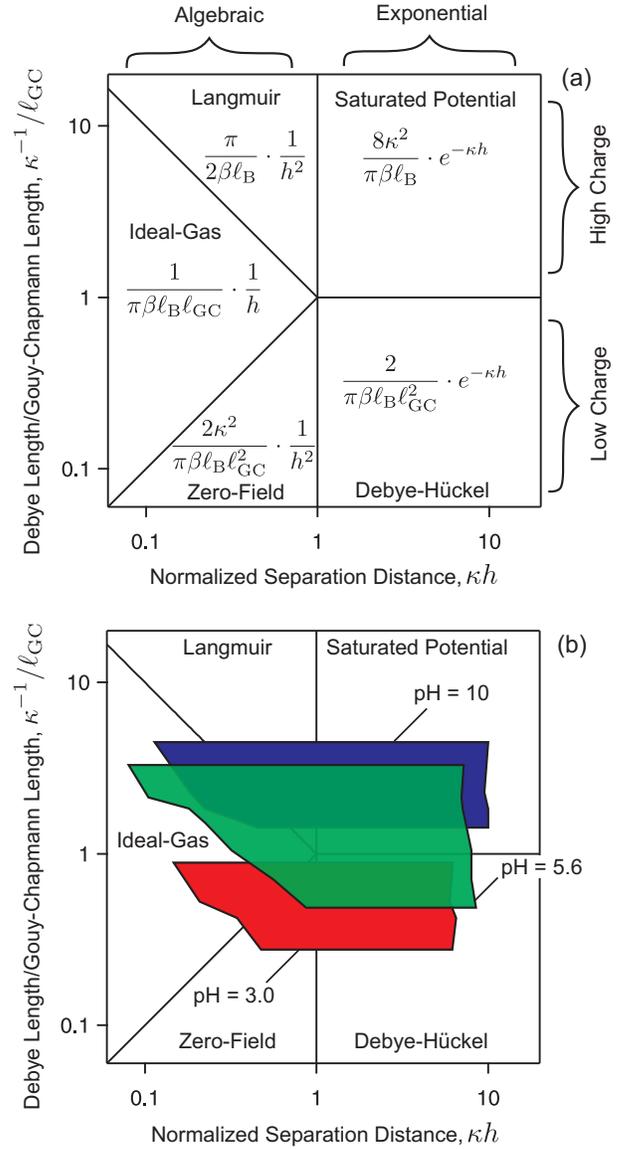}
\caption{Approximation validity map for interaction of charged plates across symmetric electrolytes. (a) Map showing the relevant regimes with equations.  (b) Experimentally accessible ranges for silica in KCl solutions at pH 3.0, pH 5.6, and pH 10 are marked with different colors.}
\label{fig:maps}
\end{figure}

Let us finally give an overview at what conditions the presented approximations are valid for symmetric electrolytes. Approximation validity maps are shown in Fig.~\ref{fig:maps}. Similar map was presented in \cite{Markovich2016a} and here we add the ranges where experimental interactions between silica colloids were measured.
In Fig.~\ref{fig:maps}a, we present the areas of validity for different approximations with the corresponding equations. The $y$-axis represents Debye length divided by the Gouy--Chapman length and the values are increasing with increasing surface charge density and decreasing salt concentration. On the $x$-axis, the separation distance relative to the Debye length is given. The map can be divided into 5 regions. Note that in reality the borders between the regions are not sharp lines as drawn here, but are more fuzzy. The map can be divided into small separation ($\kappa h < 1$) and large separation distances ($\kappa h > 1$), where algebraic and exponential approximations are dominant, respectively. At very small distances and intermediate charge, the ideal-gas equation (\ref{eq:ideal-gas}) is applicable. Note that the ideal-gas law is only valid for constant charge surfaces. At intermediate distances and for highly charged particles, the Langmuir equation \eq~(\ref{eq:pressure-lang}) is accurate. The range of validity can be extended to lower charges by using the corrected Langmuir approximation \eq~(\ref{eq:pressure-lang-corr}). At low charge and intermediate distance, the zero-field approximation \eq~(\ref{eq:zero-field2}) is applicable. At distances $\kappa h > 1$, the pressure profiles become exponential. Here we can approximate the low charged case with the Debye--H\"uckel superposition approximation, while for the highly charged surfaces one has to replace the diffuse-layer potential in the Debye--H\"uckel approximation with the effective saturated potential \cite{Elimelech2013}.

In Fig.~\ref{fig:maps}b we highlight the regions where the measurements between silica colloids were done in the present paper. The position of the range, where the force measurements are located in the map, changes depending on the salt concentration and pH of the solutions. The AFM measurements can be done reliably down to $\sim 1$~nm separation distances, while the limit for large separations depends on the magnitude of the force, which in turn depends on the charge of the surfaces. Therefore, the range where the measurements can be performed is moving down and to the right with increasing concentration, by increasing $\kappa$. On the other hand, increasing the pH shifts the ranges up due to increasing magnitude of the surface charge, which decreases $\lgc$. For all the pH conditions studied, the ideal-gas law is retrieved at low concentrations and sufficiently small separations for CC surfaces, which is consistent with the data presented in Fig.~\ref{fig:force-zero-field}. For pH 10 and pH 3, the experimental data falls into the high and low charge regimes, respectively. Therefore, the Langmuir approximation is most accurate at pH 10 for intermediate distances, while the zero-field approximation works best at pH 3. At pH 5.6 none of the $1/h^2$ approximations is very accurate since the surface charge density is in this case in the intermediate regime.

\section{Conclusions}

Algebraic double-layer interactions have been demonstrated experimentally for simple 1:1 salts. This type of force is also present for systems containing highly-charged coions, where they come from the counterion-only dominated regime. Both strongly overlapping double-layer (zero-field) and counterion-only (Langmuir) regimes recover the inverse square distance dependence of disjoining pressures, albeit with different pre-factor. The former is applicable for weakly charged surfaces and is harder to detect experimentally, while the latter is true for highly charged surfaces. Independent of the surface charge, at very small separation distances the $1/h$ ideal-gas limit is recovered for non-regulating constant charge surfaces. In the other limit, at large separation distances, the Debye--H\"uckel dependence is observed and pressures follow the conventional exponential profile. The assumption of exponential double-layer forces is therefore correct only at large separations, but at small separations the profiles are algebraic. The latter regime is also important in experimental situations, for example for colloidal aggregation in the presence of multivalent coions, or for systems containing like-charged polyelectrolytes.

\section*{Associated Content}
\noindent {\bf Supporting Information}\\
The Supporting Information is available free of charge on the ACS Publications website.\\
\\
Experimental force profiles for all pH conditions and salt levels studied. Table with parameters extracted from fitting the experimental forces.

\section*{Author Information}
\noindent {\bf Corresponding Authors}\\
\\
$^*$E-mail: \texttt{m.vis@tue.nl}\\
$^\dagger$E-mail: \texttt{gregor.trefalt@unige.ch}\\
\\
{\bf Notes}\\
The authors declare no competing financial interest.

\section*{Acknowledgments}
This research was supported by the Swiss National Science Foundation through grant 162420 and the University of Geneva. The authors are thankful to Michal Borkovec for providing access to the instruments in his laboratory and to Plinio Maroni for the help with AFM measurements. MV acknowledges the Netherlands Organisation for Scientific Research (NWO) for a Veni grant (no.~722.017.005).

\bibliography{algebraicDoubleLayerForces.bib}
\bibliographystyle{achemso}

\end{document}